\documentclass[nonacm, sigconf]{acmart}

\usepackage[english]{babel}
\usepackage[utf8]{inputenc}
\usepackage{tikz-cd}
\usepackage{lipsum}
\usepackage{xspace}
\usepackage{tcolorbox}
\usepackage{wasysym}
\usepackage{multirow}
\usepackage{tcolorbox}

\usepackage{graphicx}
\usepackage{subcaption}

\usepackage{array}
\usepackage{booktabs}
\usepackage{multirow}
\usepackage{newfloat}
\usepackage{setspace}
\usepackage{tikz-cd}
\usepackage{lipsum}

\usepackage{hyperref}
\usepackage{array}                  
\usepackage{verbatim}
\newcolumntype{L}[1]{>{\raggedright\let\newline\\\arraybackslash\hspace{0pt}}m{#1}}
\newcolumntype{C}[1]{>{\centering\let\newline\\\arraybackslash\hspace{0pt}}m{#1}}
\newcolumntype{R}[1]{>{\raggedleft\let\newline\\\arraybackslash\hspace{0pt}}m{#1}}
\newcolumntype{J}[1]{>{\let\newline\\\arraybackslash\hspace{0pt}}m{#1}}

\AtBeginDocument{%
  \providecommand\BibTeX{{%
    \normalfont B\kern-0.5em{\scshape i\kern-0.25em b}\kern-0.8em\TeX}}}

\begin{document}



\title{Understanding the Role of Large Language Models in Software Engineering: Evidence from an Industry Survey}

\author{Vítor Mateus de Brito}
\email{vitorbrito@edu.unisinos.br}
\affiliation{%
  \institution{University of Vale do Rio dos Sinos}
  \city{São Leopoldo}
  \state{Rio Grande do Sul}
  \country{Brazil}
}

\author{Kleinner Farias}
\email{kleinnerfarias@unisinos.br}
\affiliation{%
  \institution{University of Vale do Rio dos Sinos}
  \city{São Leopoldo}
  \state{Rio Grande do Sul}
  \country{Brazil}
}

\begin{abstract}
The rapid advancement of Large Language Models (LLMs) is reshaping software engineering by profoundly influencing coding, documentation, and system maintenance practices. As these tools become deeply embedded in developers’ daily workflows, understanding how they are used has become essential. This paper reports an empirical study of LLM adoption in software engineering, based on a survey of 46 industry professionals with diverse educational backgrounds and levels of experience. The results reveal positive perceptions of LLMs, particularly regarding faster resolution of technical questions, improved documentation support, and enhanced source code standardization. However, respondents also expressed concerns about cognitive dependence, security risks, and the potential erosion of technical autonomy. These findings underscore the need for critical and supervised use of LLM-based tools. By grounding the discussion in empirical evidence from industry practice, this study bridges the gap between academic discourse and real-world software development. The results provide actionable insights for developers and researchers seeking to adopt and evolve LLM-based technologies in a more effective, responsible, and secure manner, while also motivating future research on their cognitive, ethical, and organizational implications.
\end{abstract}

\keywords{Large Language Models; LLMs; Survey; Industry}

\maketitle

\section{Introdução}
O desenvolvimento de software contemporâneo ocorre em um cenário marcado por crescente complexidade, dinamicidade tecnológica e alta exigência por agilidade na entrega de soluções \cite{cabane2024impact,urdangarin2021mon4aware,rubert2022effects}. Nesse contexto, ferramentas baseadas em inteligência artificial, especialmente os Large Language Models (LLMs), têm se destacado por sua capacidade de auxiliar engenheiros de software em tarefas como geração e revisão de código, documentação, testes automatizados e suporte à modelagem e compreensão de sistemas. Apesar do crescente volume de pesquisas sobre LLMs, a literatura ainda carece de estudos empíricos que analisem como esses modelos são efetivamente empregados na prática profissional e quais impactos são percebidos por desenvolvedores em seu cotidiano.

Esses modelos, treinados em grandes volumes de dados textuais e de código, demonstram propriedades emergentes que os tornam aptos a desempenhar funções complexas que anteriormente exigiam raciocínio humano \cite{fan2023llms}. No entanto, como ressaltado por Hadi \textit{et al.} \cite{hadi2024llmsurvey}, a utilização de LLMs impõe desafios substanciais, incluindo a possibilidade de respostas incorretas (\textit{hallucinations}), falta de interpretabilidade e riscos associados à segurança e privacidade dos dados. Além disso, Zhang \textit{et al.} \cite{zhang2024survey} destacam que, embora os LLMs apresentem grande potencial para apoiar o ciclo de vida do software, sua eficácia ainda depende fortemente do contexto de uso, da forma de integração ao ambiente de desenvolvimento e do grau de especialização do modelo para tarefas específicas da engenharia de software.

Para sanar essa lacuna, este trabalho apresenta uma investigação empírica sobre o uso de LLMs na engenharia de software, conduzida por meio do método survey. O estudo contou com a participação de 46 profissionais atuantes na indústria e na academia, abrangendo diferentes níveis de experiência e formações. O questionário buscou compreender percepções, vantagens, desvantagens, riscos e desafios associados à adoção dessas ferramentas, a fim de oferecer uma visão atualizada e baseada em evidências sobre o papel dos LLMs no desenvolvimento de software. O principal diferencial deste trabalho reside em sua abordagem centrada na experiência prática dos profissionais. A contribuição da pesquisa se manifesta na oferta de um panorama empírico que auxilia pesquisadores a direcionar estudos futuros e profissionais da indústria a adotar essas tecnologias de forma mais crítica e eficiente. Ao refletir sobre percepções reais de uso, o estudo aproxima o debate acadêmico das demandas concretas do mercado de software.

Este artigo está organizado da seguinte forma: a Seção \ref{sec:background} apresenta a fundamentação teórica sobre LLMs e seu papel na engenharia de software, a Seção \ref{sec:related_works} discute os trabalhos relacionados e as lacunas identificadas, a Seção \ref{sec:metodologia} descreve a metodologia empregada e os procedimentos de coleta e análise de dados, a Seção \ref{sec:resultados} apresenta os resultados e suas interpretações e por fim, a Seção \ref{sec:conclusions} traz as conclusões e direções para pesquisas futuras.

\section{Fundamentação Teórica}
\label{sec:background}

Esta seção apresenta os fundamentos teóricos que embasam o estudo, organizando os conceitos necessários para compreender tanto o funcionamento dos Large Language Models (LLMs) (Seção \ref{fundamentacao-llms}), sua aplicação na Engenharia de Software (Seção \ref{fundamentacao-llms-eng-software}) e o método de pesquisa survey (Seção \ref{fundamentacao-survey}).

\subsection{\textit{LLMs - Large Language Models}}
\label{fundamentacao-llms}

Os \textit{Large Language Models (LLMs)}, ou grandes modelos de linguagem, constituem uma das principais inovações recentes no campo da inteligência artificial, especialmente na área de processamento de linguagem natural (PLN). Tais modelos são desenvolvidos com base em redes neurais profundas, sendo treinados sobre vastos conjuntos de dados textuais para aprender padrões, semântica e estruturas da linguagem humana e, em alguns casos, de linguagens de programação.

Segundo \cite{zheng2024code}, os LLMs vêm diminuindo a distância e diferença entre as linguagens humanas e as linguagens de máquina, devido à sua expressiva capacidade de compreender e gerar tanto textos em linguagem natural quanto em linguagens de programação. Em outras palavras, esses modelos apresentam a habilidade de interpretar comandos e perguntas formulados por pessoas, bem como de produzir códigos de computador ou analisar instruções escritas em diferentes linguagens formais. Dessa forma, os LLMs atuam como uma ponte entre a comunicação humana e as exigências computacionais, promovendo uma interação mais fluida e acessível entre usuários e sistemas digitais. Essa característica é especialmente relevante no contexto de aplicações em que a colaboração entre humanos e máquinas se torna essencial, como na assistência à programação, na tradução automática.

Em termos práticos, os LLMs servem para automatizar e aprimorar tarefas que requerem compreensão e produção textual, proporcionando respostas, resumos ou até mesmo códigos a partir de comandos em linguagem natural. Destaca-se que, no domínio da engenharia de software, surgiram variantes especializadas denominadas \textit{Code LLMs}, voltadas especificamente para tarefas como geração de código, sumarização de trechos de programação, detecção de vulnerabilidades e tradução de linguagens de programação

\subsection{LLMs na Engenharia de Software}
\label{fundamentacao-llms-eng-software}

A aplicação de Large Language Models (LLMs) na Engenharia de Software representa uma transformação significativa e recente no cenário do desenvolvimento de sistemas. Tais modelos, originalmente projetados para tarefas de processamento de linguagem natural, vêm sendo progressivamente adaptados para lidar com atividades que envolvem linguagens de programação e artefatos técnicos, impulsionando tanto a automação quanto a inovação em diferentes estágios do ciclo de vida do software.

De acordo com \cite{fan2023llms}, as propriedades emergentes dos LLMs têm viabilizado sua utilização em uma ampla gama de tarefas na Engenharia de Software, incluindo geração e completamento de código, design, elicitação e análise de requisitos, reparo automatizado, refatoração, otimização de desempenho, documentação e análise de software. Ferramentas baseadas em LLMs, como o GitHub Copilot e o Cursor, já demonstram impacto prático, auxiliando desenvolvedores por meio de sugestões automáticas, geração de trechos de código e explicações detalhadas de funcionalidades.

As principais aplicações dos modelos de linguagem de larga escala (LLMs) na engenharia de software abrangem várias etapas do desenvolvimento. Entre elas, destaca-se a \textbf{geração e completamento de código}, na qual os LLMs sugerem ou produzem automaticamente trechos de código a partir de descrições em linguagem natural. Também se destaca o \textbf{reparo automatizado e a depuração}, com modelos capazes de identificar falhas e propor correções de forma autônoma.

No \textbf{teste de software}, os LLMs auxiliam na criação de casos de teste e na detecção de comportamentos inesperados, aumentando a robustez das aplicações. Já na \textbf{refatoração e otimização}, contribuem para aprimorar sistemas legados e propor melhorias de desempenho. Por fim, na \textbf{documentação e explicação de código}, esses modelos geram documentação automática e explicações de trechos complexos, facilitando a comunicação e a manutenção dos sistemas.

\subsection{Método de Pesquisa \textit{Survey}}
\label{fundamentacao-survey}

O método de pesquisa \textit{survey} configura-se como uma abordagem quantitativa amplamente utilizada nas ciências sociais e aplicadas, cujo principal objetivo é a obtenção de dados acerca das características, comportamentos, opiniões ou atitudes de um grupo específico de pessoas. Tal método se destaca pela sua capacidade de coletar informações padronizadas junto a uma amostra representativa, permitindo a análise de fenômenos em larga escala e a generalização dos resultados para uma população mais ampla.

Segundo \cite{freitas2000survey}, o \textit{survey} é particularmente apropriado quando o intuito da pesquisa é responder a questões como “o quê?”, “por que?”, “como?” e “quanto?”, ou seja, quando o foco está em compreender “o que está acontecendo”, bem como “como e por que isso está acontecendo”. Essa flexibilidade torna o método útil tanto para investigações iniciais quanto para estudos mais avançados e explicativos.

Ainda conforme \cite{freitas2000survey}, o método \textit{survey} pode assumir diferentes propósitos, destacando-se três categorias principais. A \textbf{explanatória} visa testar teorias e relações causais entre variáveis, buscando compreender os mecanismos subjacentes aos fenômenos estudados. A \textbf{exploratória} tem como objetivo familiarizar-se com determinado tópico, identificar conceitos iniciais e definir quais aspectos devem ser medidos e de que maneira isso deve ocorrer. Por fim, a \textbf{descritiva} procura mapear e identificar situações, eventos, atitudes ou opiniões manifestas em uma população, oferecendo um retrato detalhado dos fenômenos investigados.

A escolha do método \textit{survey}, portanto, justifica-se especialmente em contextos onde há necessidade de sistematizar informações a partir de um grande número de respondentes, possibilitando tanto análises descritivas quanto inferenciais dos dados coletados \cite{freitas2000survey}. Dessa forma, o \textit{survey} constitui-se em uma ferramenta valiosa para pesquisas que buscam compreender padrões, tendências e relações em ambientes organizacionais, educacionais ou tecnológicos.

\textbf{Coleta de Dados}. A etapa de coleta de dados constitui uma fase fundamental no delineamento de pesquisas do tipo \textit{survey}, visto que a qualidade e a fidedignidade dos dados obtidos impactam diretamente a robustez dos resultados e das conclusões do estudo. Conforme destacado por \cite{wohlin2012experimentation}, os dois meios mais comuns de coleta de dados em pesquisas \textit{survey} são os questionários e as entrevistas. Os questionários, por sua vez, podem ser disponibilizados em formato impresso ou eletrônico, seja por meio de e-mails ou páginas web, facilitando o alcance de um maior número de participantes em diferentes contextos e localidades.

No âmbito desta pesquisa, optou-se pela utilização do método \textit{survey}, sendo a coleta de dados realizada por meio de questionários estruturados. Essa escolha se justifica pela possibilidade de obtenção de uma amostra representativa da população de interesse, uma vez que o formato de questionário permite a seleção de uma ampla gama de participantes de maneira ágil e eficiente. Além disso, o emprego do questionário como instrumento de coleta de dados proporciona a padronização das respostas, conferindo maior comparabilidade e consistência aos dados coletados \cite{wohlin2012experimentation}.

Dessa forma, a análise das respostas fornecidas pelos participantes possibilita uma compreensão mais abrangente das opiniões, percepções e comportamentos do público-alvo, alinhando-se aos objetivos do estudo. Portanto, a escolha pelo uso de questionários eletrônicos em \textit{surveys} evidencia-se como uma estratégia metodológica eficiente para a coleta de dados em pesquisas contemporâneas.

\section{Trabalhos Relacionados}
\label{sec:related_works}

Foram realizadas buscas em repositórios digitais de trabalhos científicos como o Google Scholar, em seguidas os trabalhos foram analisados e comparados.

\subsection{Análise dos Trabalhos Relacionados}
\label{subsec:related_works_analysis}

\textbf{Zheng \textit{et al.} (2024) \cite{zheng2024code}}. O artigo apresenta uma revisão sistemática sobre Large Language Models (LLMs) aplicados a tarefas de engenharia de software, com foco especial em modelos desenvolvidos ou ajustados especificamente para código (Code LLMs). Foram coletados e analisados 149 trabalhos relevantes, a partir de uma busca em quatro grandes bases de dados e repositórios (dblp, Google Scholar, GitHub e arXiv), empregando métodos rigorosos de triagem e análise de dados, incluindo card sorting e snowballing. Os resultados apontam que, em geral, Code LLMs superam LLMs generalistas em tarefas específicas de engenharia de software, especialmente quando ambos possuem tamanhos de parâmetros similares e os modelos são finamente ajustados para tarefas do domínio. Entretanto, os LLMs generalistas de última geração, como o GPT-4, ainda mostram desempenho competitivo, principalmente em tarefas variadas. O artigo também discute a influência do tamanho do modelo, os impactos do fine-tuning, e apresenta uma análise comparativa abrangente de desempenho em benchmarks como HumanEval, CSN e outros. Por fim, ressalta a necessidade de mais investigações sobre métricas, benchmarks e o desenvolvimento de Code LLMs voltados a desafios específicos da engenharia de software, sugerindo que a área ainda possui muitos desafios em aberto para pesquisa futura.

\textbf{Jiang \textit{et al.} (2024) \cite{jiang2024codegen}}. O artigo apresenta uma revisão abrangente e atualizada sobre o uso de Large Language Models (LLMs) para geração de código, com foco especial nos avanços recentes e aplicações práticas. Os autores propõem uma taxonomia detalhada que abrange desde curadoria de dados, técnicas de pré-treinamento, avaliação de desempenho e implicações éticas até o impacto ambiental e usos em ambientes reais. O estudo também realiza uma análise histórica da evolução dos LLMs voltados à geração de código, incluindo comparações empíricas entre modelos populares em benchmarks como HumanEval, MBPP e BigCodeBench. Por fim, o trabalho identifica desafios críticos — como a lacuna entre pesquisa e prática — e oportunidades futuras, destacando o potencial dos LLMs para revolucionar o desenvolvimento de software.

\textbf{Zhang \textit{et al.} (2024a) \cite{zhang2024survey}}. Apresenta um levantamento abrangente sobre o uso de modelos de linguagem de grande escala (LLMs) aplicados à engenharia de software. O artigo categoriza os modelos segundo sua origem (indústria, academia e comunidades open source) e analisa suas capacidades em tarefas como geração de código, geração de testes, sumarização, tradução de código e reparo de vulnerabilidades. Os autores também discutem benchmarks, avanços recentes e desafios, como confiança, alucinação e avaliação da qualidade das saídas. Conclui-se que, apesar do progresso significativo, ainda existem lacunas consideráveis para melhorar a precisão, robustez e aplicabilidade prática dos LLMs na engenharia de software.

\textbf{Zhang \textit{et al.} (2024b) \cite{zhang2024apr}}. O artigo apresenta uma revisão sistemática sobre o uso de Large Language Models (LLMs) na tarefa de Automated Program Repair (APR). Foram analisados 127 estudos publicados entre 2020 e 2024, com o objetivo de compreender as principais técnicas, modelos utilizados, cenários de reparo abordados e os desafios enfrentados na integração de LLMs com APR. O trabalho categoriza os LLMs utilizados em três grupos (encoder-only, encoder-decoder e decoder-only), analisa abordagens de adaptação como fine-tuning, few-shot e zero-shot, e identifica os principais tipos de bugs tratados, com destaque para erros semânticos e vulnerabilidades de segurança. Os autores também discutem os fatores que influenciam a eficácia das soluções propostas, como formatos de entrada, uso de benchmarks e a questão da sobreposição de dados de treino (data leakage). Conclui-se que, embora os LLMs tenham promovido avanços notáveis na área, ainda há diversos desafios a serem enfrentados, como altos custos computacionais, falta de estudos empíricos com desenvolvedores e necessidade de benchmarks mais robustos e livres de vazamento de dados.

\textbf{Shi \textit{et al.} (2025)\cite{shi2023greenllm}}. O artigo apresenta uma visão estratégica para o futuro dos Large Language Models aplicados à Engenharia de Software (LLM4SE), com ênfase na eficiência computacional e na sustentabilidade ambiental. Os autores defendem a urgência de se desenvolver modelos que consumam menos recursos computacionais, memória, energia, água e emitam menos carbono, democratizando o acesso a essas ferramentas para pequenas empresas e desenvolvedores individuais. Por meio de uma revisão sistemática da literatura, são identificadas técnicas promissoras organizadas em quatro eixos: centradas em dados, modelos, sistemas e programas. O trabalho propõe ainda uma agenda de pesquisa até 2030, incluindo benchmarks dedicados, métodos de treinamento mais eficientes, novas técnicas de compressão e aceleração de inferência, bem como otimizações nos programas subjacentes aos modelos. A conclusão reforça que o futuro da engenharia de software exige LLMs não apenas eficazes, mas também acessíveis e ecologicamente responsáveis.

\textbf{Hadi \textit{et al.} (2024) \cite{hadi2024llmsurvey}}. O artigo apresenta um levantamento abrangente sobre os Modelos de Linguagem de Grande Escala (LLMs), abordando sua evolução, funcionamento, principais aplicações, desafios e limitações. São discutidos aspectos técnicos como arquitetura, pré-processamento e estratégias de prompting, além de aplicações em áreas como medicina, educação, engenharia e direito. O estudo também explora preocupações éticas, consumo energético, viés algorítmico e a necessidade de regulamentações. Por fim, aponta direções futuras para o uso responsável dos LLMs, destacando questões abertas e sugerindo diretrizes para pesquisas futuras.

\textbf{Fan \textit{et al.} (2023) \cite{fan2023llms}}. O artigo realiza uma ampla revisão da aplicação de Large Language Models (LLMs) na Engenharia de Software, abordando desde tarefas como geração de código, testes, documentação e manutenção até problemas abertos e desafios científicos. Os autores destacam o papel emergente dos LLMs e suas propriedades não determinísticas, chamando atenção para os riscos de alucinações e a importância de técnicas híbridas que combinem LLMs com métodos tradicionais da engenharia de software. O trabalho também propõe direções futuras, como o aprimoramento de avaliações científicas, uso de explicações geradas por LLMs, testes automatizados e engenharia de prompts, defendendo que a colaboração entre LLMs e engenheiros humanos será essencial para a evolução segura e eficaz da disciplina.

\textbf{Hou \textit{et al.} (2023) \cite{hou2023srl}}. Apresenta uma revisão sistemática da literatura sobre o uso de Large Language Models (LLMs) em Engenharia de Software, abrangendo 229 estudos publicados entre 2020 e 2023. A revisão classificou os trabalhos segundo tarefas de software, técnicas de aplicação e tipos de contribuição, identificando um crescimento significativo no uso de LLMs em tarefas como geração de código, sumarização, busca e correção automática de bugs. O estudo revela que a maioria dos trabalhos ainda está em fase preliminar, com muitos artigos sendo publicados em repositórios abertos como o arXiv. Conclui-se que o campo está em rápida expansão, com desafios abertos relacionados à avaliação científica rigorosa, engenharia de prompt, e cobertura desigual entre as subáreas da engenharia de software.

\textbf{Liu \textit{et al.} (2024) \cite{liu2024qual}}. O artigo realiza uma análise empírica sistemática sobre a qualidade do código gerado pelo ChatGPT (baseado em GPT-3.5), utilizando um conjunto de 2.033 tarefas de programação do LeetCode e avaliando 4.066 programas gerados em Python e Java. Os autores identificam que, embora a maioria dos códigos gerados seja funcionalmente correta, uma parcela significativa apresenta problemas de estilo, manutenção, erros de execução e saídas incorretas. A pesquisa também explora a capacidade de autocorreção do ChatGPT mediante feedbacks simples e feedbacks detalhados com base em ferramentas de análise estática e erros de execução, demonstrando que o modelo é capaz de resolver de 20\% a 60\% dos problemas. Ainda assim, novas falhas podem ser introduzidas durante o processo de correção. O estudo conclui que, embora promissor, o uso de LLMs como ChatGPT para geração de código requer cuidado quanto à confiabilidade e qualidade do código gerado, sugerindo o uso de feedback contextualizado e engenharia de prompts como caminhos para mitigar os problemas.

\textbf{Yang \textit{et al.} (2024) \cite{yang2024robust}} conduzem uma revisão sistemática da literatura com o objetivo de identificar e analisar propriedades não-funcionais em modelos de linguagem de larga escala voltados para código (LLM4Code), indo além da tradicional métrica de acurácia. A partir de 146 estudos relevantes, os autores destacam sete propriedades cruciais: robustez, segurança, privacidade, explicabilidade, eficiência e usabilidade. A revisão revela lacunas substanciais na literatura atual, como a fragilidade dos modelos frente a pequenas perturbações, vulnerabilidades a ataques de envenenamento de dados, riscos de vazamento de informações sensíveis, falta de explicações compreensíveis para usuários finais, alto custo computacional e impactos ambíguos na produtividade. Os autores propõem uma agenda de pesquisa baseada em três perspectivas — centrada em dados, em humanos e em sistemas — para guiar o aprimoramento das propriedades não-funcionais dos LLM4Code em contextos reais de uso.

\subsection{Análise Comparativa e Identificação de Lacunas de Pesquisa}
\label{trabalhos-relacionados-comparativo}

Os critérios empregados para a análise comparativa — \textbf{metodologia utilizada}, \textbf{população ou corpus analisado}, \textbf{principais resultados e descobertas}, e \textbf{ferramentas ou benchmarks empregados} — foram selecionados com base em parâmetros amplamente validados em estudos anteriores sobre o uso de LLMs na Engenharia de Software. Esses critérios permitem uma avaliação sistemática e comparável entre diferentes abordagens, possibilitando identificar tendências, limitações e oportunidades de avanço no campo.

A partir da análise dos trabalhos apresentados na Tabela~\ref{tab:tabela_llms_relacionados}, que sintetiza as principais pesquisas revisadas, foi possível observar um predomínio de estudos voltados a revisões sistemáticas da literatura e análises empíricas baseadas em benchmarks técnicos. Embora tais investigações tenham contribuído significativamente para o entendimento das capacidades e limitações dos modelos de linguagem, ainda persistem lacunas importantes que merecem exploração em pesquisas futuras.

Em especial, destacam-se três lacunas principais. Primeiramente, há uma \textbf{ausência de estudos centrados na experiência e percepção dos profissionais de engenharia de software}, uma vez que a maioria das pesquisas adota uma perspectiva predominantemente técnica ou automatizada. Em segundo lugar, observa-se a \textbf{escassez de análises que considerem o impacto prático dos LLMs na rotina de trabalho}, incluindo aspectos como produtividade percebida, adaptação de fluxos e barreiras à adoção. Por fim, verifica-se uma \textbf{carência de dados qualitativos e quantitativos baseados em experiências reais de uso}, que possam complementar métricas laboratoriais e oferecer uma visão mais contextualizada do papel dos LLMs na prática profissional.

Diante dessas lacunas, o presente estudo propõe uma abordagem empírica que busca compreender de forma direta o impacto dos LLMs na rotina de profissionais da área.

\begin{table*}[!ht].
    \centering
    \caption{Análise comparativa entre os trabalhos selecionados}
\begin{tabular}{|p{2.3cm}|p{2.3cm}|p{2.8cm}|p{4cm}|p{3.5cm}|}
\hline
\footnotesize
\textbf{Trabalho} & \textbf{Metodologia} & \textbf{População / Corpus} & \textbf{Principais Resultados} & \textbf{Ferramentas/Benchmarks} \\
\hline
Zheng \textit{et al.} (2024) \cite{zheng2024code} & Revisão Sistemática da Literatura (SLR) & 134 artigos (2021–2023) & Code LLMs superam LLMs gerais em tarefas de código & HumanEval, MBPP, CodeXGLUE \\
\hline
Jiang \textit{et al.} (2024) \cite{jiang2024codegen} & SLR & 235 artigos (2018 – 2024) & Avanços recentes em LLMs para geração de código; desafios em robustez e avaliação & HumanEval, MBPP, BigCodeBench \\
\hline
Zhang \textit{et al.} (2024a) \cite{zhang2024survey} & SLR & 1009 estudos (2020 – 2024) & LLMs são aplicados em 112 tarefas de SE; desafios de avaliação e integração & CodeXGLUE, HumanEval, diversos \\
\hline
Zhang \textit{et al.} (2024b) \cite{zhang2024apr} & SLR & 127 artigos (2020 – 2024) & LLMs impulsionam o avanço do APR; há limitações em custos, dados e avaliação & Defects4J, QuixBugs, BFP, TFix, HumanEval-Java \\
\hline
Shi \textit{et al.} (2023) \cite{shi2023greenllm} & Survey / Vision & Revisão de literatura e propostas & Desafios de eficiência e sustentabilidade; roadmap para LLMs mais verdes e acessíveis & CodeXGLUE, EffiBench, outros \\
\hline
Hadi \textit{et al.} (2024) \cite{hadi2024llmsurvey} & Revisão narrativa & Diversos estudos e aplicações & Avanços, aplicações, desafios e limitações dos LLMs em vários domínios & ChatGPT, Bard, Llama, Claude, frameworks diversos \\
\hline
Fan \textit{et al.} (2023) \cite{fan2023llms} & Survey narrativa & Diversos estudos e tendências recentes & LLMs impactam várias tarefas de SE; desafios de avaliação e integração & HumanEval, LeetCode, SWE-bench, PapersWithCode \\
\hline
Hou \textit{et al.} (2023) \cite{hou2023srl} & SLR (revisão sistemática) & 229 artigos (2017 – 2023) & LLMs avançam em várias tarefas de SE, principalmente geração de código; desafios em avaliação, generalização e integração & HumanEval, MBPP, Defects4J, benchmarks diversos \\
\hline
Liu \textit{et al.} (2024) \cite{liu2024qual} & Estudo empírico & 2.033 tarefas (LeetCode); 4.066 códigos (Java/Python) & ChatGPT gera código funcional, mas com problemas de qualidade; feedback ajuda, mas há limitações & LeetCode, Pylint, Flake8, PMD, Checkstyle, dataset público \\
\hline
Yang \textit{et al.} (2024) \cite{yang2024robust} & Revisão sistemática & 146 artigos (2019 – 2024) & Avanços, mas desafios em robustez, segurança, explicabilidade, eficiência e usabilidade & CodeBERT, Copilot, datasets públicos, benchmarks diversos \\
\hline
\end{tabular}
    \label{tab:tabela_llms_relacionados}
\end{table*}

\section{Metodologia}
\label{sec:metodologia}

Esta seção descreve a metodologia adotada para conduzir o presente estudo. A Seção \ref{subsec:metodologia-objetivo} apresenta o objetivo geral da pesquisa e a questão de pesquisa investigada. A Seção \ref{subsec:metodologia-processo} detalha o processo metodológico seguido, com ênfase nas etapas de planejamento, coleta e análise de dados. Por fim, a Seção \ref{subsec:metodologia-questionario} apresenta os aspectos específicos relacionados ao questionário aplicado. A condução do estudo seguiu diretrizes para pesquisa empírica em Engenharia de Software propostas por \cite{wohlin2012experimentation}, além de trabalhos recentes sobre o uso de LLMs no contexto da engenharia de software. Além disso, este trabalho adotou práticas e orientações previamente validadas e publicadas em estudos empíricos anteriores \cite{junior2022use,junior2021survey,farias2018uml,farias2013analyzing}.

\subsection{Objetivo e Questão de Pesquisa}
\label{subsec:metodologia-objetivo}

O objetivo deste estudo é investigar o uso de modelos de linguagem de larga escala (LLMs) no contexto da engenharia de software, com foco nas atividades práticas do desenvolvimento, como escrita, revisão e compreensão de código. A pesquisa busca compreender como profissionais da área estão utilizando essas ferramentas em seu dia a dia, quais benefícios têm percebido, quais desafios enfrentam, e qual o impacto percebido na produtividade, qualidade do código e nos processos de tomada de decisão técnica. A principal questão de pesquisa é apresentada a seguir:


\begin{itemize}
    
    \item \textbf{QP:} Quais são os impactos percebidos do uso de LLMs no cotidiano de profissionais de engenharia de software? 
    
\end{itemize}

\subsection{Processo Metodológico}
\label{subsec:metodologia-processo}

\begin{figure*}[h!]
\centering
\includegraphics[scale=.5]{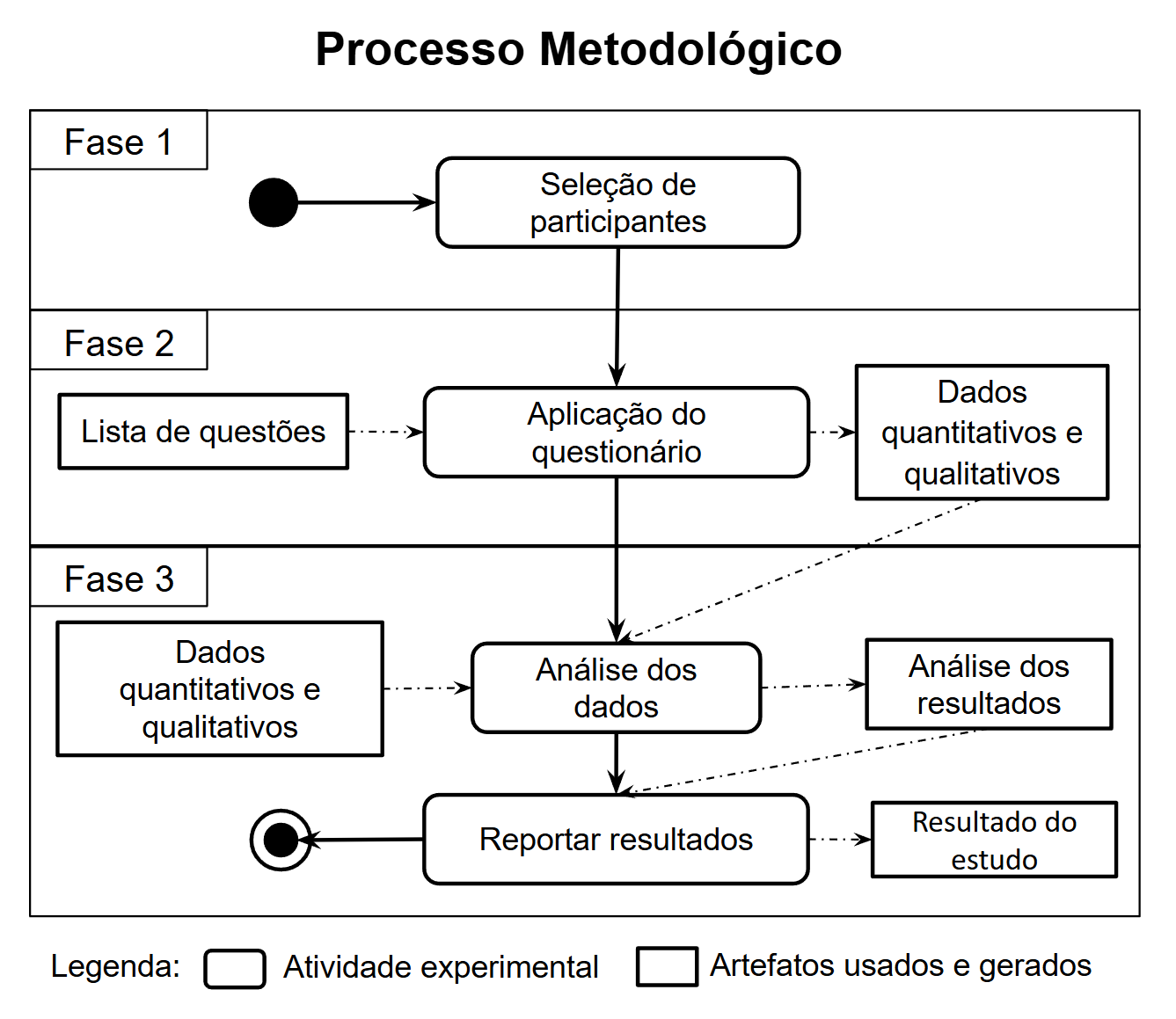}
\caption{Processo experimental}
\label{fig:processo-metodologico}
\end{figure*}

Figura \ref{fig:processo-metodologico} apresenta o processo metodológico adotado. Ele é baseado em estudos experimentais validados e publicados na literatura \cite{farias2016empirical,farias2014effects,farias2015evaluating}.

\textbf{Fase 1: Seleção de participantes.} Os participantes foram selecionados com base nos seguintes critérios: atuação profissional na área de desenvolvimento ou engenharia de software, familiaridade com ferramentas baseadas em LLMs (como GitHub Copilot, ChatGPT, Gemini, DeepSeek, entre outras), e experiência prática em projetos da indústria. O público-alvo incluiu desenvolvedores de software, engenheiros de software, arquitetos e líderes técnicos, de diferentes níveis de senioridade, atuando principalmente no Brasil. Esta delimitação visa garantir a relevância dos dados para o contexto nacional e profissional.

\textbf{Fase 2: Aplicação do questionário.} Foi desenvolvido um questionário estruturado contendo questões fechadas e abertas. O instrumento foi validado preliminarmente por meio de revisão por pares e enviado aos participantes por meio de formulário eletrônico. O formulário foi enviado direto por e-mail aos participantes selecionados. A coleta teve como foco dados quantitativos e qualitativos sobre a frequência de uso, tipos de tarefas auxiliadas, percepções de produtividade, desafios enfrentados e aspectos éticos ou organizacionais envolvidos no uso de LLMs.

\textbf{Fase 3: Análise dos dados.} Os dados obtidos por meio do questionário foram analisados inicialmente de forma descritiva e estatística, com o apoio de ferramentas de visualização e tabulação. A análise qualitativa das respostas abertas foi conduzida por meio da técnica de codificação aberta da Grounded Theory \cite{glaser2017discovery}, permitindo a identificação de categorias emergentes relacionadas ao impacto e uso dos LLMs. As evidências obtidas foram utilizadas para responder à questão de pesquisa e fundamentar a discussão dos resultados.

\subsection{Questionário}
\label{subsec:metodologia-questionario}

O questionário foi elaborado com o objetivo de obter uma visão ampla sobre o uso de LLMs na rotina de profissionais da engenharia de software. A ferramenta utilizada para a aplicação foi o Google Forms, considerando sua praticidade, acessibilidade e amplo alcance. A Tabela~\ref{tab:questoes-llms} apresenta os pontos trabalhadores e sua motivação. As perguntas foram organizadas em blocos temáticos: perfil do participante, experiências com LLMs, percepção de vantagens e desvantagens, impacto percebido no processo de desenvolvimento e questões éticas. Todas as respostas foram anônimas e os dados coletados serão utilizados exclusivamente para fins acadêmicos. A estrutura do questionário buscou equilibrar questões fechadas (com múltiplas escolhas e escalas Likert) com perguntas abertas, de modo a captar tanto aspectos objetivos quanto subjetivos da experiência dos participantes.

\begin{table*}[!ht].
    \centering
    \caption{Questões e motivações sobre o uso de LLMs em desenvolvimento de software}
\begin{tabular}{|p{0.5cm}|p{8cm}|p{8cm}|}
\hline
\footnotesize
\textbf{N°} & \textbf{Questão} & \textbf{Motivação} \\
\hline
Q1 & O uso de LLMs em equipes distribuídas/remotas é fundamental para manter a agilidade e eficiência do desenvolvimento. & Verificar se apoiam a colaboração em equipes distribuídas/remotas. \\ 
\hline
Q2 & O uso de LLMs pode contribuir para a identificação e antecipação de riscos arquiteturais e limitações tecnológicas em projetos de software. & Identificar se ajudam na antecipação de riscos arquiteturais e tecnológicos. \\ 
\hline
Q3 & LLMs promovem a disseminação de conhecimento sobre novas tecnologias e boas práticas dentro da equipe. & Examinar se promovem a disseminação de conhecimento e boas práticas. \\ 
\hline
Q4 & LLMs auxiliam no onboarding e treinamento de novos(as) desenvolvedores(as) em equipes de software. & Avaliar se facilitam o onboarding e o treinamento de novos(as) desenvolvedores(as). \\ 
\hline
Q5 & O uso de LLMs pode dificultar a revisão e manutenção do código devido à geração automática de trechos não compreendidos totalmente pelo desenvolvedor. & Analisar se a geração automática pode dificultar manutenção e revisão. \\ 
\hline
Q6 & A integração de LLMs com ferramentas de desenvolvimento (IDEs, repositórios, pipelines) é simples e eficiente. & Avaliar a simplicidade e eficiência de integração com ferramentas de desenvolvimento. \\ 
\hline
Q7 & LLMs contribuem para a padronização do código-fonte (boas práticas, padrões de projeto etc.). & Avaliar se contribuem para padronização e qualidade do código. \\ 
\hline
Q8 & O uso de LLMs pode introduzir riscos de segurança devido à geração de código não auditado. & Avaliar riscos de segurança com geração de código não auditado. \\ 
\hline
Q9 & LLMs contribuem para a detecção e correção mais rápida de bugs. & Medir se contribuem para detecção e correção mais rápida de bugs. \\ 
\hline
Q10 & O uso de LLMs pode induzir à dependência excessiva das sugestões automáticas, prejudicando o desenvolvimento das habilidades técnicas dos(as) desenvolvedores(as). & Explorar riscos de dependência excessiva que comprometa habilidades técnicas. \\
\hline
Q11 & LLMs auxiliam na elaboração, atualização e revisão da documentação técnica de sistemas. & Verificar se auxiliam na elaboração e atualização da documentação técnica. \\ 
\hline
Q12 & O uso de LLMs reduz o tempo de resolução de dúvidas técnicas no cotidiano do desenvolvimento. & Medir o impacto na redução do tempo de resolução de dúvidas técnicas. \\  
\hline
Q13 & LLMs são eficazes na geração automática de testes de software (unitários, integração etc.). & Investigar se são eficazes na geração automática de testes. \\
\hline
Q14 & O uso de LLMs facilita a compreensão e manutenção de código legado. & Verificar se LLMs reduzem a complexidade na leitura e manutenção de código legado. \\
\hline
Q15 & O uso de LLMs auxilia na redução do retrabalho causado por requisitos mal definidos ou modificados ao longo do desenvolvimento. & Identificar se auxiliam na mitigação de retrabalho por requisitos mal definidos. \\ 
\hline
\end{tabular}
\label{tab:questoes-llms}
\end{table*}

\section{Resultados}
\label{sec:resultados}

Esta seção apresenta os resultados obtidos a partir da análise das respostas ao questionário aplicado (conforme descrito na Seção 4.3). Gráficos foram utilizados para sintetizar e analisar as percepções dos 46 participantes sobre o uso de modelos de linguagem de larga escala (LLMs) na engenharia de software.

\subsection{Análise de Perfil dos Participantes}
\label{subsec:analysis}

A Tabela~\ref{tab:demographic-data} apresenta o perfil dos participantes da pesquisa, incluindo informações sobre faixa etária, formação acadêmica, nível de escolaridade, tempo de estudo, cargo ocupado, tempo na função e experiência com desenvolvimento de software.

\begin{table}[h!]
\scriptsize
\centering
\caption{Dados de perfil dos participantes.}
\label{tab:demographic-data}
\begin{tabular}{lccc}
\hline
\textbf{Característica} & \textbf{Resposta} & \textbf{\#} & \textbf{\%} \\ 
\hline

\multicolumn{1}{l}{Faixa etária} 
    & 18–25 anos & 9 & 19,6\% \\
    & 26–30 anos & 12 & 26,1\% \\
    & 31–35 anos & 13 & 28,3\% \\
    & 36–40 anos & 5 & 10,9\% \\
    & 41–45 anos & 2 & 4,3\% \\
    & > 45 anos & 4 & 8,7\% \\
    & Não respondeu & 1 & 2,2\% \\
\hline

\multicolumn{1}{l}{Graduação} 
    & Ciência da Computação & 10 & 21,7\% \\
    & Análise de Sistemas & 12 & 26,1\% \\
    & Sistemas de Informação & 9 & 19,6\% \\
    & Sistemas para Internet & 2 & 4,3\% \\
    & Engenharia da Computação & 2 & 4,3\% \\
    & Outros & 11 & 24,0\% \\
\hline

\multicolumn{1}{l}{Escolaridade} 
    & Superior Incompleto & 11 & 23,9\% \\
    & Superior Completo & 15 & 32,6\% \\
    & Pós-graduação & 10 & 21,7\% \\
    & Mestrado & 7 & 15,2\% \\
    & Outros & 3 & 6,6\% \\
\hline

\multicolumn{1}{l}{Tempo de estudo formal} 
    & < 3 anos & 9 & 19,6\% \\
    & 3–4 anos & 9 & 19,6\% \\
    & 5–6 anos & 11 & 23,9\% \\
    & 7–8 anos & 9 & 19,6\% \\
    & > 8 anos & 8 & 17,4\% \\
\hline

\multicolumn{1}{l}{Cargo/posição}
    & Engenheiro de Software & 23 & 50,0\% \\
    & Especialista & 4 & 8,7\% \\
    & Analista & 2 & 4,3\% \\
    & Estagiário & 5 & 10,9\% \\
    & Outros & 12 & 26,1\% \\
\hline

\multicolumn{1}{l}{Tempo nesse cargo/posição}
    & < 3 anos & 24 & 52,2\% \\
    & 3–4 anos & 11 & 23,9\% \\
    & 5–6 anos & 4 & 8,7\% \\
    & 7–8 anos & 0 & 0,0\% \\
    & > 8 anos & 7 & 15,2\% \\
\hline

\multicolumn{1}{l}{Tempo de experiência com}
    & < 3 anos & 13 & 28,3\% \\
\multicolumn{1}{l}{desenvolvimento de software}
    & 3–4 anos & 10 & 21,7\% \\
    & 5–6 anos & 12 & 26,1\% \\
    & 7–8 anos & 1 & 2,2\% \\
    & > 8 anos & 10 & 21,7\% \\
\hline
\end{tabular}
\end{table}

\textbf{Faixa etária.}
A maioria dos participantes (54,4\%) está concentrada entre 26 e 35 anos, sendo 26,1\% na faixa de 26 a 30 anos e 28,3\% na faixa de 31 a 35 anos. Esse dado evidencia um público predominantemente adulto jovem. Além disso, 19,6\% dos respondentes estão entre 18 e 25 anos, o que indica a presença de profissionais em início de carreira. Observa-se também que 23,9\% dos participantes possuem mais de 36 anos, incluindo 8,7\% acima dos 45 anos. Essa diversidade etária contribui para uma visão ampla sobre o tema estudado, englobando tanto perspectivas de profissionais mais jovens quanto mais velhos.

\textbf{Graduação.}
A formação acadêmica dos participantes é predominantemente voltada para cursos da área de Tecnologia da Informação, com destaque para Análise de Sistemas (26,1\%), Ciência da Computação (21,7\%) e Sistemas de Informação (19,6\%). Esses três cursos somam 67,4\% das respostas, indicando um público fortemente técnico. Além disso, 8,6\% possuem graduação em Engenharia da Computação ou Sistemas para Internet, e 24\% indicaram outras formações.

\textbf{Escolaridade.}
A maioria dos participantes já possui formação superior completa (32,6\%) ou está em níveis mais avançados, como pós-graduação (21,7\%) e mestrado (15,2\%). Em conjunto, 69,5\% possuem pelo menos o nível superior completo, o que reforça o caráter especializado da amostra. Por outro lado, 23,9\% ainda estão cursando o ensino superior, o que é coerente com a presença de profissionais mais jovens apontada anteriormente. O equilíbrio entre profissionais formados e em formação permite observar diferentes estágios de maturidade acadêmica e profissional.

\textbf{Tempo de estudo formal.}
Os resultados indicam uma distribuição relativamente equilibrada entre as faixas de tempo de estudo. As categorias de 5 a 6 anos (23,9\%), menos de 3 anos (19,6\%), 3 a 4 anos (19,6\%) e 7 a 8 anos (19,6\%) possuem proporções semelhantes, o que mostra trajetórias educacionais diversas. Além disso, 17,4\% relataram mais de 8 anos de estudo formal, o que pode estar relacionado à continuidade acadêmica (como pós-graduação) ou à retomada de cursos superiores em diferentes momentos da carreira. Essa pluralidade temporal de formação reflete diferentes ritmos de desenvolvimento profissional e acadêmico dentro do grupo.

\textbf{Cargo/posição.}
Metade dos participantes (50\%) atua como engenheiro de software, o que demonstra uma amostra fortemente concentrada em atividades diretamente ligadas ao desenvolvimento de sistemas. Outras funções como especialista (8,7\%), analista (4,3\%) e estagiário (10,9\%) também aparecem, evidenciando a presença de profissionais em distintos níveis hierárquicos e estágios de carreira. A categoria “Outros” (26,1\%) sugere diversidade de cargos correlatos, como líderes técnicos, funções de gestão técnica, entre outros. Esse panorama é particularmente relevante para a pesquisa, pois contempla profissionais com experiências tanto operacionais quanto estratégicas no ciclo de vida do software.

\textbf{Tempo nesse cargo/posição.}
Mais da metade dos respondentes (52,2\%) ocupa o cargo atual há menos de 3 anos, o que sugere um grupo com relativa mobilidade profissional ou em fase de consolidação de carreira. A faixa de 3 a 4 anos (23,9\%) indica uma quantidade expressiva de profissionais em nível intermediário, enquanto 15,2\% afirmaram estar há mais de 8 anos na mesma posição, representando o grupo mais experiente. Essa diversidade temporal é importante para compreender como diferentes níveis de maturidade profissional influenciam a percepção sobre as práticas de engenharia de software.

\textbf{Tempo de experiência com desenvolvimento de software.}
Os dados apontam uma amostra heterogênea em termos de experiência. Embora 28,3\% tenham menos de 3 anos na área, 60,5\% possuem mais de 4 anos de experiência, o que inclui 26,1\% entre 5 a 6 anos e 21,7\% com mais de 8 anos. Essa combinação de perfis — desde iniciantes até profissionais seniores — contribui significativamente para a qualidade das respostas, permitindo observar percepções que refletem diferentes graus de envolvimento prático e teórico no desenvolvimento de software.

\subsection{Impacto}
\label{subsec:impact}

\begin{figure*}[]
    \centering
    \includegraphics[scale=0.6]{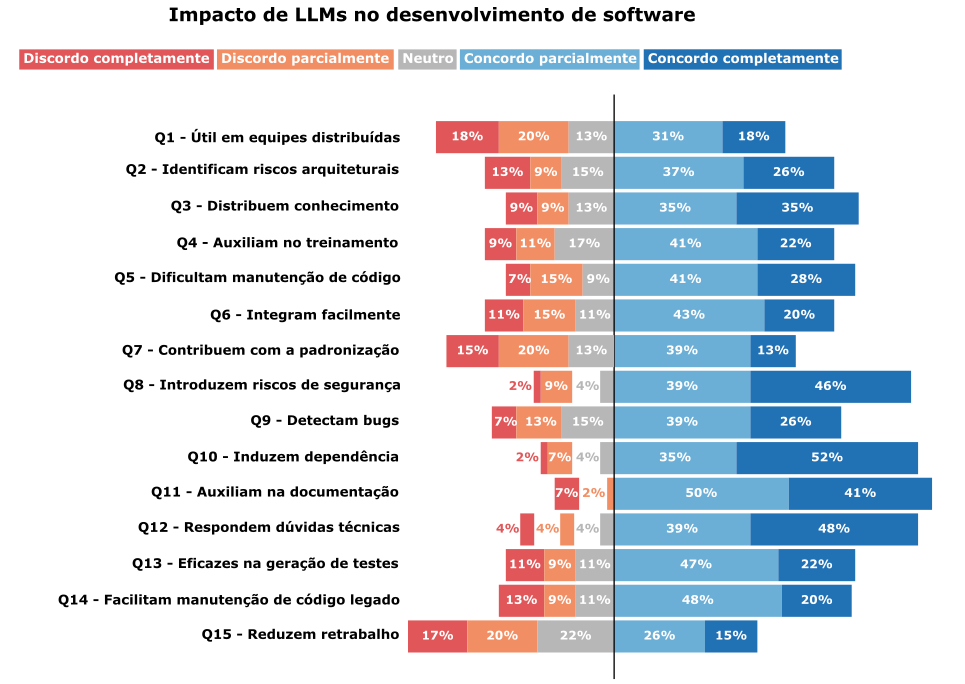}
    \caption{Gráfico Likert}
    \label{fig:graph-likert}
\end{figure*}

A Figura \ref{fig:graph-likert} apresenta os resultados obtidos sobre o impacto de LLMs no desenvolvimento de software, que investigaram o grau de concordância dos participantes em relação a diferentes afirmações sobre os efeitos do uso de modelos de linguagem em atividades de engenharia de software. Foram analisadas 46 respostas, distribuídas em uma escala Likert de 5 pontos, variando de “Discordo completamente” a “Concordo completamente”.

De forma geral, os resultados indicam uma percepção predominantemente positiva quanto ao impacto dos LLMs no contexto do desenvolvimento de software. Em diversos itens, as respostas se concentraram nas opções de concordância, evidenciando que a maioria dos participantes reconhece contribuições significativas dessas ferramentas para a produtividade, a aprendizagem e a qualidade do código.

Entre as afirmações com maior grau de concordância, destaca-se que 87\% dos participantes (somando “Concordo parcialmente” e “Concordo completamente”) consideram que o uso de LLMs auxilia na redução do tempo de resolução de dúvidas técnicas no cotidiano do desenvolvimento. De forma semelhante, 69\% dos respondentes concordam que os LLMs são eficazes na geração automatizada de testes (unitários, integração etc.), enquanto 68\% afirmam que essas ferramentas auxiliam na compreensão e manutenção de código legado. Esses resultados reforçam a visão de que os LLMs têm se consolidado como instrumentos de apoio operacional, capazes de otimizar tarefas rotineiras e reduzir o esforço cognitivo dos desenvolvedores.

Outro ponto relevante é o reconhecimento do papel dos LLMs na documentação e padronização de código. Aproximadamente 91\% dos participantes concordaram que os modelos contribuem para a elaboração e atualização de documentação técnica e padronização de boas práticas de código-fonte, indicando que essas ferramentas são percebidas como assistentes de produtividade não apenas na geração de código, mas também na criação e atualização de documentações dos sistemas.

Em contrapartida, alguns riscos e limitações também foram reconhecidos pelos participantes. Cerca de 87\% concordaram que o uso de LLMs pode induzir à dependência excessiva, prejudicando o desenvolvimento do pensamento crítico. Ademais, 85\% dos respondentes reconheceram que os LLMs podem introduzir riscos de segurança devido à geração automática de código não auditado, e 69\% concordaram que essas ferramentas podem dificultar a revisão e manutenção de código em virtude de trechos gerados sem pleno entendimento por parte dos desenvolvedores.

Além disso, o uso de LLMs foi amplamente associado à disseminação de conhecimento e de boas práticas técnicas. Aproximadamente 70\% dos participantes concordaram que os modelos auxiliam na aprendizagem de novas tecnologias e 63\% que concordam que ela facilita o onboarding de novos(as) desenvolvedores(as) nas equipes, o que sugere um papel formativo importante dessas ferramentas, especialmente em contextos de integração de novos membros ou aprendizado técnico. Também foi observada concordância majoritária (63\%) quanto à ideia de que os LLMs contribuem para identificar riscos arquiteturais e limitações tecnológicas, indicando que essas ferramentas estão sendo utilizadas não apenas para geração de código, mas também como instrumentos de análise e suporte à decisão técnica.

Por outro lado, afirmações mais neutras ou divergentes apareceram em itens relacionados à integração dos LLMs com ferramentas de desenvolvimento e à redução de retrabalho causado por requisitos mal definidos. Nesses casos, observou-se maior dispersão das respostas e percentuais mais equilibrados entre as opções “Neutro” e “Concordo parcialmente”, sugerindo que, embora o impacto positivo seja reconhecido, sua efetividade plena depende de fatores contextuais, como o nível de integração da ferramenta com o ecossistema técnico da equipe e a maturidade dos processos de desenvolvimento.

A análise da Figura \ref{fig:graph-likert} demonstra que os participantes percebem o impacto dos LLMs no desenvolvimento de software de maneira amplamente positiva, especialmente no que se refere à agilidade, documentação, testes e manutenção de código. Entretanto, os resultados também apontam alertas importantes relacionados à dependência cognitiva e à segurança do código gerado. Assim, o uso dos LLMs deve ser compreendido como um recurso de apoio técnico e formativo, que potencializa o desempenho das equipes quando empregado com criticidade, supervisão humana e práticas de validação contínua.

\subsection{Vantagens e Desvantagens}
\label{subsec:advantages-disadvantages}

\begin{figure*}[h!]
    \centering
    \includegraphics[scale=0.6]{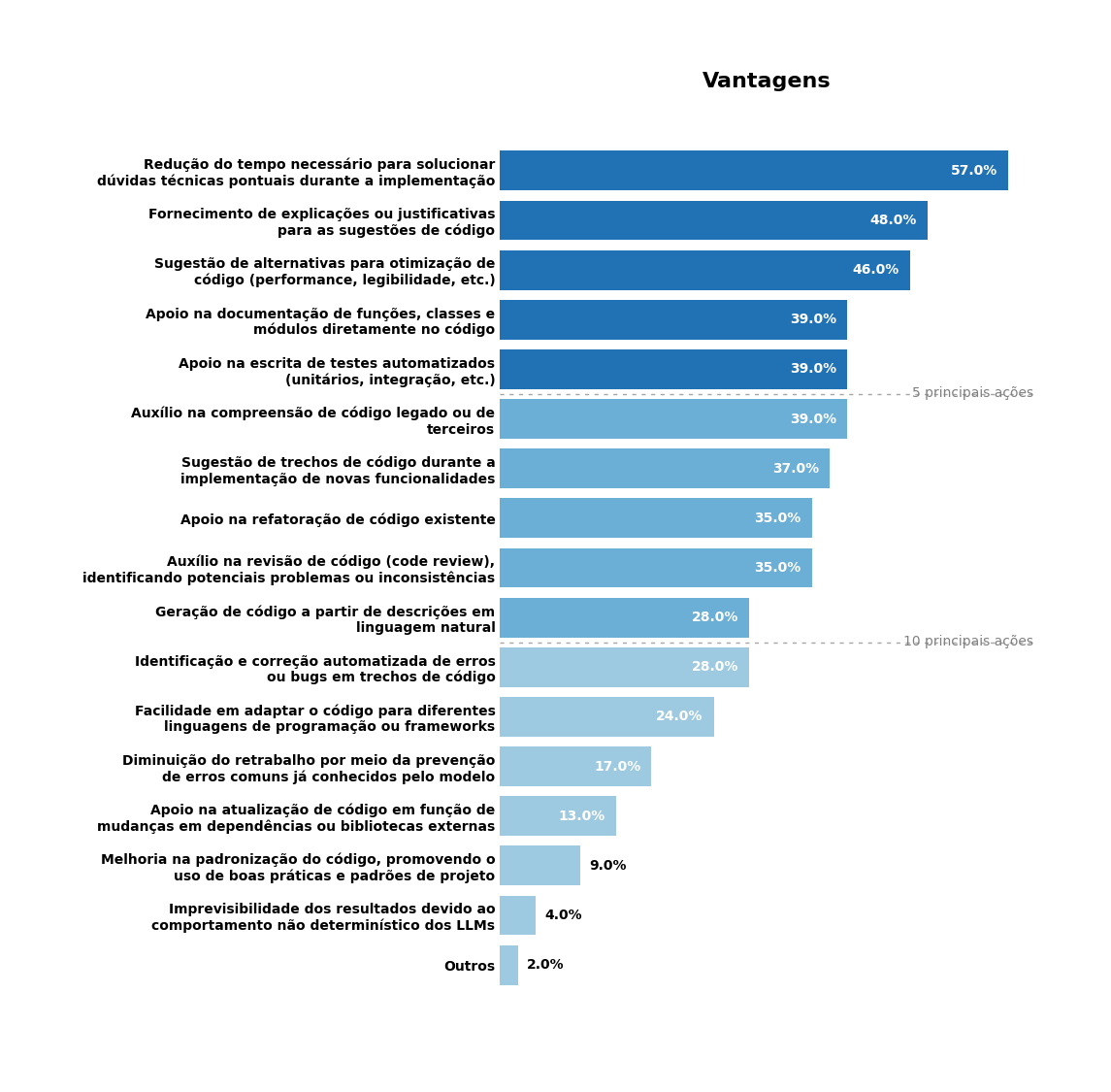}
    \caption{Vantagens}
    \label{fig:graph-vantagens}
\end{figure*}

A Figura \ref{fig:graph-vantagens} apresenta os dados coletados sobre a variável vantagens do uso de LLMs. As respostas estão organizadas em ordem decrescente conforme a frequência das escolhas. De modo geral, observa-se uma distribuição relativamente equilibrada entre as opções apresentadas, embora algumas vantagens se destaquem de forma expressiva.

A principal vantagem reportada pelos participantes foi a redução do tempo necessário para solucionar dúvidas técnicas pontuais durante a implementação, com 57\% das respostas. Esse resultado evidencia a função prática e imediata dos modelos de linguagem no suporte ao desenvolvimento de software, especialmente em tarefas de depuração e esclarecimento conceitual. Em seguida, aparecem o fornecimento de explicações ou justificativas para as sugestões de código (48\%) e a sugestão de alternativas para otimização do código (46\%), apontando que os participantes valorizam não apenas a geração de código em si, mas também a capacidade explicativa e o potencial de aprimoramento técnico oferecido pelos LLMs.

Entre as cinco principais vantagens, também se destacam o apoio na documentação de funções, classes e módulos diretamente no código (39\%) e o auxílio na escrita de testes automatizados (39\%). Esses resultados reforçam a percepção de que o uso de LLMs pode contribuir para a melhoria da qualidade estrutural e da manutenibilidade do software, ao incentivar práticas de engenharia consolidadas, como documentação e testes. Além disso, o auxílio na compreensão de código legado ou de terceiros, que figura empatado na quinta posição (39\%), demonstra o papel dos LLMs na leitura e interpretação de bases de código complexas, muitas vezes consideradas tarefas de alta carga cognitiva para desenvolvedores humanos.

No conjunto das dez principais vantagens, nota-se a presença de aspectos ligados tanto à produtividade quanto à qualidade do código. A sugestão de trechos de código durante a implementação de novas funcionalidades (37\%) e o apoio à refatoração (35\%) exemplificam o potencial dos LLMs como assistentes durante o fluxo de desenvolvimento. Já o auxílio em revisões de código (code review) (35\%) reforça o uso dessas ferramentas como mecanismos de verificação adicional, capazes de identificar possíveis inconsistências e propor melhorias de legibilidade.

As demais vantagens listadas apresentam menores frequências, mas ainda oferecem insights relevantes. Por exemplo, a geração de código a partir de descrições em linguagem natural (28\%) e a identificação e correção automatizada de erros (28\%) ilustram a confiança dos participantes na capacidade dos modelos de transformar requisitos textuais em implementações práticas e detectar falhas de maneira proativa. Apesar disso, vantagens mais específicas, como a padronização do código (9\%) ou o apoio em mudanças de dependências externas (13\%), aparecem com menor destaque, sugerindo que os participantes percebem maior utilidade dos LLMs em atividades diretamente relacionadas à codificação e suporte imediato ao desenvolvedor.

A análise evidencia que as principais vantagens percebidas no uso de LLMs estão associadas à agilidade no desenvolvimento, melhoria da compreensão técnica e apoio à qualidade do código. As ações mais citadas refletem o uso dos modelos como ferramentas assistivas de produtividade e documentação, mais do que como substitutos autônomos de desenvolvedores. Esses resultados fortalecem a visão de que os LLMs têm potencial para atuar como facilitadores do trabalho técnico, ampliando a eficiência e a qualidade do processo de desenvolvimento sem eliminar o papel crítico do julgamento humano.

\begin{figure*}[h!]
    \centering
    \includegraphics[scale=0.468]{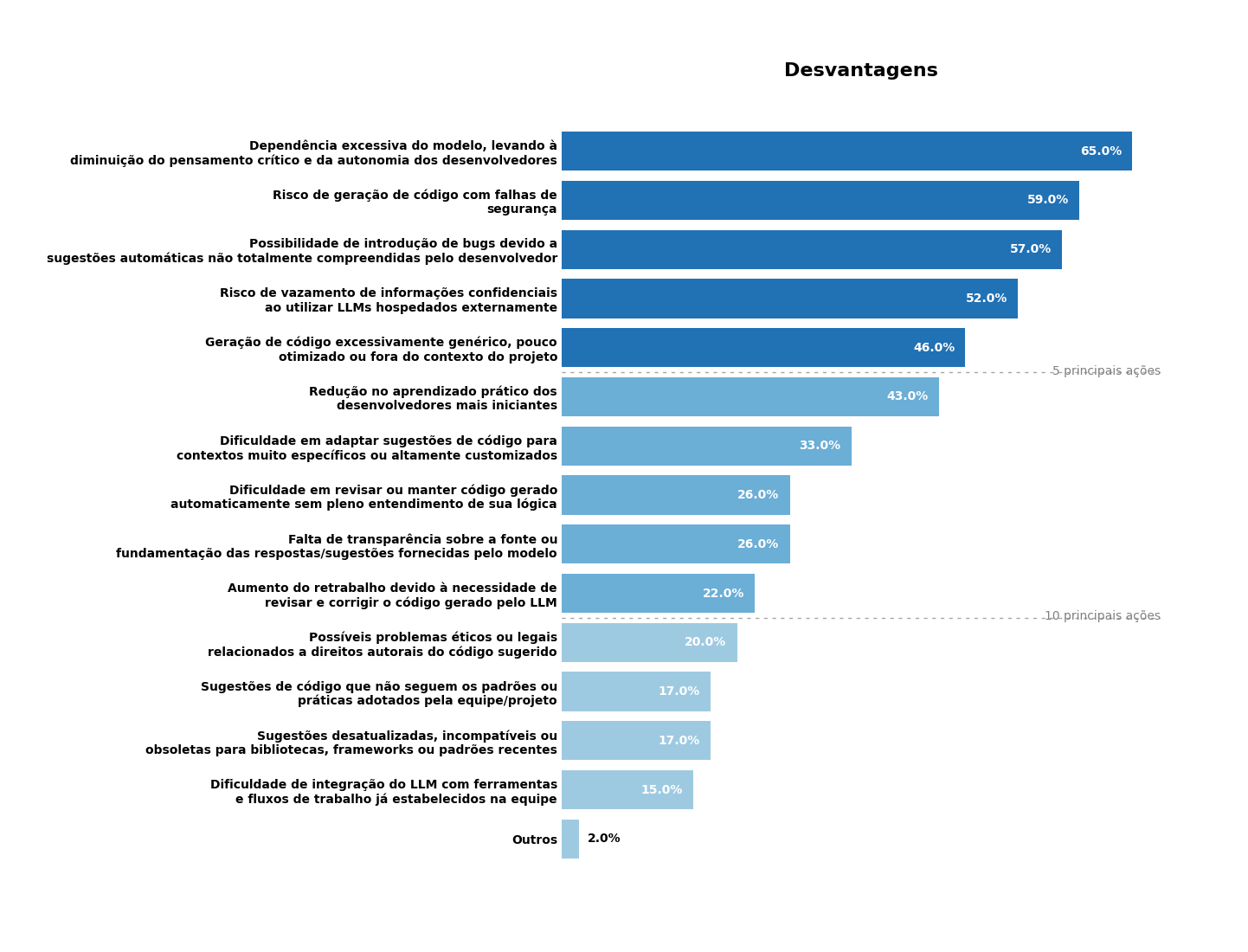}
    \caption{Desvantagens}
    \label{fig:graph-desvantagens}
\end{figure*}

A Figura \ref{fig:graph-desvantagens} apresenta os dados coletados referentes à variável desvantagens do uso de LLMs. Respostas organizadas em ordem decrescente de frequência. A distribuição dos resultados demonstra que há uma clara concentração das respostas nas primeiras opções, indicando consenso entre os participantes sobre os principais riscos e limitações associados ao uso de modelos de linguagem no desenvolvimento de software.

A dependência excessiva do modelo, levando à diminuição do pensamento crítico e da autonomia dos desenvolvedores, foi a desvantagem mais citada, mencionada por 65\% dos participantes. Esse resultado evidencia uma preocupação recorrente na literatura e entre profissionais da área: a possibilidade de que a adoção contínua de ferramentas baseadas em LLMs reduza a capacidade reflexiva e analítica dos desenvolvedores, tornando-os excessivamente dependentes das sugestões automáticas.

Em segundo e terceiro lugar aparecem o risco de geração de código com falhas de segurança (59\%) e a possibilidade de introdução de bugs devido a sugestões automáticas não totalmente compreendidas (57\%). Esses resultados reforçam que, embora os LLMs ofereçam ganhos de produtividade, sua utilização sem verificação humana pode comprometer a robustez e a segurança do sistema.

Entre as cinco principais desvantagens, observa-se ainda o risco de vazamento de informações confidenciais ao utilizar LLMs hospedados externamente (52\%) e a geração de código genérico, pouco otimizado ou fora do contexto do projeto (46\%). Ambas refletem preocupações com privacidade e adequação contextual. O primeiro ponto é especialmente relevante em ambientes corporativos, onde a exposição de dados sensíveis a modelos hospedados em nuvem pode representar violações de compliance e de propriedade intelectual. O segundo ponto, por sua vez, demonstra limitações técnicas dos modelos em compreender integralmente o contexto arquitetural e as especificidades de cada projeto.

As desvantagens subsequentes — redução no aprendizado prático dos desenvolvedores mais iniciantes (43\%) e dificuldade em adaptar sugestões de código a contextos específicos (33\%) — complementam esse panorama, sugerindo que o uso intensivo de LLMs pode impactar negativamente o processo de aprendizagem e a autonomia técnica, especialmente em equipes em formação. Além disso, a dificuldade em revisar ou manter código gerado automaticamente (26\%) e a falta de transparência sobre as fontes e justificativas das respostas dos modelos (26\%) reforçam preocupações quanto à rastreabilidade e interpretabilidade das soluções produzidas. 

As demais desvantagens apresentaram menores percentuais, mas ainda são relevantes sob o ponto de vista organizacional. O aumento do retrabalho (22\%), problemas éticos e legais relacionados à autoria de código (20\%) e sugestões de código que não seguem os padrões da equipe (17\%) apontam para implicações práticas no fluxo de trabalho e na manutenção da consistência técnica do projeto. Além disso, sugestões desatualizadas ou incompatíveis com bibliotecas recentes (17\%) e dificuldades de integração do LLM com ferramentas já utilizadas pela equipe (15\%) evidenciam limitações operacionais que podem comprometer a adoção sustentável dessas tecnologias em ambientes corporativos.

A análise revela que as principais desvantagens percebidas no uso de LLMs concentram-se em três eixos principais: dependência cognitiva, riscos técnicos e preocupações éticas e de segurança. Os participantes reconhecem que, embora os modelos de linguagem acelerem o desenvolvimento, também introduzem desafios significativos em termos de confiança, privacidade e qualidade do código. Esses resultados reforçam a necessidade de políticas internas, boas práticas e capacitação contínua para garantir que o uso de LLMs ocorra de forma crítica, segura e alinhada às demandas arquiteturais e organizacionais de cada contexto.

\subsection{Discussão Adicional}
\label{subsec:impact}

A questão aberta \textbf{“Qual sua opinião sobre o uso de LLMs na Engenharia de Software?”} buscou compreender, de forma descritiva, as percepções individuais dos participantes acerca do papel, benefícios e limitações dos modelos de linguagem no contexto do desenvolvimento de software. As respostas foram analisadas qualitativamente, considerando recorrências temáticas, nuances de posicionamento e o tom geral das percepções expressas.

De modo geral, a maioria dos participantes apresentou uma visão predominantemente positiva, reconhecendo os LLMs como ferramentas poderosas, úteis e complementares ao trabalho humano, capazes de aumentar a produtividade, agilizar a resolução de dúvidas técnicas, auxiliar na escrita de código, testes e documentação e melhorar a eficiência de processos repetitivos. Expressões como “ferramenta poderosa”, “ótimo recurso” e “excelente ferramenta para produtividade” foram frequentes, indicando uma ampla aceitação do uso dessas tecnologias quando aplicadas de forma criteriosa e supervisionada.

Entre os principais benefícios mencionados, destacam-se:
\begin{itemize}\setlength\itemsep{0.2em}
    \item Aumento de produtividade e redução do tempo de desenvolvimento, frequentemente associados à automatização de tarefas de baixo valor cognitivo.
    \item Suporte à aprendizagem e ao desenvolvimento profissional, especialmente para consulta de dúvidas, documentação e revisão de código.
    \item Apoio à qualidade e padronização, com menções positivas à capacidade dos LLMs de auxiliar na estruturação de código, elaboração de testes e detecção de erros.
    \item Ampliação do acesso ao conhecimento técnico, atuando como um assistente de fácil consulta, comparado por alguns participantes a “um Google melhorado”.
\end{itemize}

No entanto, as respostas também revelaram preocupações significativas, refletindo uma percepção crítica e madura sobre o uso dessas ferramentas. As principais limitações e riscos apontados foram:
\begin{itemize}\setlength\itemsep{0.2em}
    \item Dependência excessiva dos modelos, especialmente entre desenvolvedores iniciantes, levando à redução do pensamento crítico e à perda de autonomia cognitiva.
    \item Risco de erros e falhas de segurança, em virtude da geração de código sem total compreensão por parte do usuário.
    \item Superficialidade no aprendizado técnico, quando o uso da ferramenta substitui o esforço de entendimento conceitual.
    \item Falta de confiabilidade das respostas e necessidade de revisão constante pelo desenvolvedor humano.
\end{itemize}

Essas preocupações se alinham diretamente aos resultados quantitativos das questões Q2 e Q10, que também destacaram, respectivamente, o risco de falhas técnicas e a dependência cognitiva como os principais desafios associados ao uso de LLMs. Muitos participantes enfatizaram a importância de uso consciente e supervisionado, reforçando que a ferramenta deve servir como assistente e não substituto do engenheiro de software. Afirmações como “ajuda muito, mas precisa ser revisado pelo desenvolvedor humano” e “ótima ferramenta, mas só se usada da maneira correta” sintetizam esse equilíbrio entre entusiasmo e prudência.

Além disso, algumas respostas apresentaram reflexões de caráter mais estratégico e institucional. Participantes mencionaram a necessidade de estruturas de suporte e conscientização por parte das empresas e instituições de ensino, de modo a garantir uso ético, responsável e sustentável dessas tecnologias. Um participante observou que o uso de LLMs é “fundamental desde de que haja um consciente, ou seja, seguido de toda uma estrutura de suporte e conscientização”, sugerindo que a integração dessas ferramentas deve ser acompanhada por políticas organizacionais claras e formação contínua.

Outros comentários destacaram a dimensão evolutiva e inevitável do uso das LLMs, ressaltando que elas representam uma tendência de mercado irreversível e um diferencial competitivo para profissionais que souberem aplicá-las adequadamente. Nesse sentido, as LLMs foram vistas não apenas como uma ferramenta técnica, mas como um marco transformacional no processo de engenharia de software, capaz de redefinir a natureza do trabalho e a formação dos futuros desenvolvedores.

A análise das respostas descritivas revela um consenso entre os participantes de que os LLMs representam uma inovação significativa e inevitável na Engenharia de Software, oferecendo ganhos expressivos de produtividade, aprendizado e qualidade, desde que seu uso seja crítico, consciente e supervisionado. A percepção geral é de complementaridade, e não de substituição, os LLMs são vistos como ferramentas assistivas que ampliam as capacidades humanas, mas que exigem responsabilidade, conhecimento técnico e discernimento para evitar dependência, erros e prejuízos à autonomia profissional.

A questão aberta \textbf{“Como você tem utilizado LLMs na sua rotina?”} teve como objetivo compreender de que forma os participantes têm incorporado o uso de modelos de linguagem em suas atividades diárias. As respostas revelam uma adoção ampla e diversificada, indicando que os LLMs já ocupam um papel significativo na rotina de desenvolvedores e estudantes de Engenharia de Software.

De modo geral, as menções mais recorrentes relacionam-se ao apoio em atividades de codificação, incluindo a geração de trechos curtos de código, sugestões de funções, explicações de sintaxe e resolução de dúvidas técnicas. Alguns participantes descreveram o LLM como uma espécie de “Google aprimorado”, utilizado principalmente para buscas rápidas e direcionadas ao contexto do código em desenvolvimento. Também é comum o uso das ferramentas para converter código entre linguagens, compreender código legado e analisar erros, o que demonstra sua utilidade tanto na aprendizagem quanto na manutenção de sistemas já existentes.

Outro grupo expressivo de respostas destacou o uso de LLMs na refatoração e revisão de código, especialmente em tarefas de padronização e melhoria de legibilidade. Há relatos de utilização dos modelos para gerar mensagens de commit, descrições de pull requests e resumos explicativos em revisões de código, indicando uma integração cada vez maior ao fluxo colaborativo de desenvolvimento. Da mesma forma, o apoio à documentação aparece como uma das funções mais valorizadas, incluindo a escrita e atualização de diagramas, relatórios técnicos e resumos de especificações (como PDD e SDD).

Também foi amplamente mencionado o uso dos LLMs para geração de testes automatizados, sobretudo em casos simples e repetitivos, permitindo ao desenvolvedor concentrar-se em tarefas mais complexas. Alguns participantes relataram empregar agentes automatizados em ciclos iterativos de geração e correção de testes, demonstrando um estágio mais avançado de integração dessas ferramentas ao processo de desenvolvimento.

Além das tarefas diretamente ligadas ao código, vários participantes relataram utilizar os LLMs para automação de atividades administrativas, como redação de e-mails formais, elaboração de atas, relatórios e resumos de documentos extensos. Em contextos pessoais e acadêmicos, também foram mencionados usos voltados à organização de estudos, planejamento de tarefas e apoio à pesquisa, reforçando o caráter multifuncional dessas ferramentas.

As respostas também evidenciam uma integração crescente dos LLMs em todo o ciclo de engenharia de software, desde a ideação até a documentação final. Essa presença transversal indica que os modelos estão sendo incorporados de forma estratégica ao fluxo de trabalho, principalmente para eliminar tarefas repetitivas e de baixo valor cognitivo, liberando tempo para que o desenvolvedor se dedique a atividades criativas, analíticas e arquiteturalmente mais complexas.

As respostas indicam que os LLMs vêm sendo amplamente utilizados como assistentes multifuncionais no cotidiano dos profissionais de Engenharia de Software, com destaque para a geração e revisão de código, automação de testes, documentação e apoio à aprendizagem. O uso é percebido como altamente benéfico, desde que acompanhado de discernimento técnico e revisão humana.

\section{Conclusão e Trabalhos Futuros}
\label{sec:conclusions}

Este trabalho realizou um survey com o objetivo de descrever e analisar como os Large Language Models (LLMs) vêm sendo utilizados na prática na engenharia de software. A pesquisa foi conduzida junto a 46 profissionais atuantes na indústria, abrangendo diferentes níveis de experiência e formações. Por meio da coleta e análise das respostas, buscou-se compreender as percepções dos participantes quanto aos benefícios, riscos e desafios relacionados à adoção dessas ferramentas, fornecendo um panorama atualizado sobre o impacto dos LLMs no cotidiano do desenvolvimento de software.

Os resultados obtidos indicam que os LLMs têm sido amplamente reconhecidos como ferramentas de apoio que promovem ganhos expressivos de produtividade, aprendizado e qualidade do código, especialmente em tarefas como documentação, testes e compreensão de código legado. Contudo, os participantes também demonstraram consciência crítica sobre suas limitações, destacando riscos de dependência cognitiva, falhas de segurança e perda de autonomia técnica. A principal contribuição deste estudo reside em oferecer uma visão empírica, baseada em evidências do uso real de LLMs na indústria, permitindo que pesquisadores compreendam as implicações práticas desses modelos e que profissionais adotem estratégias mais seguras e eficazes de integração dessas tecnologias em seus fluxos de trabalho.

Como trabalhos futuros, sugere-se a ampliação da amostra e a condução de estudos que avaliem o impacto dos LLMs ao longo do tempo em diferentes contextos organizacionais. Além disso, futuras pesquisas podem investigar comparativamente o desempenho de equipes com e sem apoio de LLMs, bem como explorar aspectos éticos, de segurança e de governança de IA aplicados ao desenvolvimento de software. Outra direção promissora consiste em integrar métodos qualitativos, como entrevistas em profundidade, a fim de captar nuances comportamentais e cognitivas associadas ao uso dessas ferramentas, contribuindo para uma compreensão ainda mais abrangente e crítica de seu papel na engenharia de software contemporânea.

\section*{Questões Éticas}
\label{sec:ethics}

A pesquisa foi conduzida em contextos onde a aprovação ética para estudos de levantamento foi considerada desnecessária devido à natureza da investigação, que envolveu participação voluntária sem quaisquer riscos potenciais ou coleta de informações sensíveis dos participantes.

\section*{Agradecimentos}

Este trabalho foi apoiado pelo Conselho Nacional de Desenvolvimento Cient\'{i}fico e Tecnol\'{o}gico (CNPq), No. 312320/2025-6

\bibliographystyle{ACM-Reference-Format}
\bibliography{references}

\break
\section*{Apêndice A: Questionários}

\begin{table*}[]
\centering
\caption{Questões de perfil dos participantes}
\begin{tabular}{|l|}
\hline
\footnotesize
\textbf{Questão} \\
\hline
\textbf{1:} Qual sua idade? \\
\hline
\textbf{2:} Qual a escolaridade? \\
\hline
\textbf{3:} Qual sua área de formação acadêmica? \\
\hline
\textbf{4:} Quanto tempo de estudo formal? \\
\hline
\textbf{5:} Quanto tempo de experiência em desenvolvimento/engenharia de software? \\
\hline
\textbf{6:} Qual seu cargo atualmente? \\
\hline
\textbf{7:} Há quanto tempo está nesse cargo/posição? \\
\hline
\textbf{8:} Quanto tempo de experiência você tem com engenharia de software? \\
\hline
\end{tabular}
\label{tab:questoes-llms1}
\end{table*}

\begin{table*}[!ht].
\centering
\caption{Questões de pesquisa investigadas neste artigo}
\begin{tabular}{|p{15cm}|}
\hline
\footnotesize
\textbf{Questões sobre o uso de LLMs em desenvolvimento de software} \\
\hline
\textbf{1:} O uso de LLMs facilita a compreensão e manutenção de código legado. \\
\hline
\textbf{2:} LLMs contribuem para a padronização do código-fonte (boas práticas, padrões de projeto etc.). \\
\hline
\textbf{3:} O uso de LLMs auxilia na redução do retrabalho causado por requisitos mal definidos ou modificados ao longo do desenvolvimento. \\
\hline
\textbf{4:} LLMs promovem a disseminação de conhecimento sobre novas tecnologias e boas práticas dentro da equipe. \\
\hline
\textbf{5:} LLMs auxiliam no onboarding e treinamento de novos(as) desenvolvedores(as) em equipes de software. \\
\hline
\textbf{6:} O uso de LLMs reduz o tempo de resolução de dúvidas técnicas no cotidiano do desenvolvimento. \\
\hline
\textbf{7:} LLMs são eficazes na geração automática de testes de software (unitários, integração etc.). \\
\hline
\textbf{8:} O uso de LLMs pode dificultar a revisão e manutenção do código devido à geração automática de trechos não compreendidos totalmente pelo desenvolvedor. \\
\hline
\textbf{9:} O uso de LLMs em equipes distribuídas/remotas é fundamental para manter a agilidade e eficiência do desenvolvimento. \\
\hline
\textbf{10:} O uso de LLMs pode introduzir riscos de segurança devido à geração de código não auditado. \\
\hline
\textbf{11:} O uso de LLMs pode contribuir para a identificação e antecipação de riscos arquiteturais e limitações tecnológicas em projetos de software. \\
\hline
\textbf{12:} LLMs contribuem para a detecção e correção mais rápida de bugs. \\
\hline
\textbf{13:} LLMs auxiliam na elaboração, atualização e revisão da documentação técnica de sistemas. \\
\hline
\textbf{14:} A integração de LLMs com ferramentas de desenvolvimento (IDEs, repositórios, pipelines) é simples e eficiente. \\
\hline
\textbf{15:} O uso de LLMs pode induzir à dependência excessiva das sugestões automáticas, prejudicando o desenvolvimento das habilidades técnicas dos(as) desenvolvedores(as). \\
\hline
\end{tabular}
\label{tab:questoes-llms2}
\end{table*}

\begin{table*}[!ht].
\centering
\caption{Vantagens percebidas no uso de LLMs em Engenharia de Software}
\begin{tabular}{|p{15cm}|}
\hline
\footnotesize
\textbf{Principais Vantagens identificadas} \\
\hline
Geração de código a partir de descrições em linguagem natural \\
\hline
Sugestão de alternativas para otimização de código (performance, legibilidade, etc.) \\
\hline
Fornecimento de explicações ou justificativas para as sugestões de código \\
\hline
Apoio na escrita de testes automatizados (unitários, integração, etc.) \\
\hline
Diminuição do retrabalho por meio da prevenção de erros comuns já conhecidos pelo modelo \\
\hline
Apoio na refatoração de código existente \\
\hline
Auxílio na compreensão de código legado ou de terceiros \\
\hline
Melhoria na padronização do código, promovendo o uso de boas práticas e padrões de projeto \\
\hline
Apoio na atualização de código em função de mudanças em dependências ou bibliotecas externas \\
\hline
Identificação e correção automatizada de erros ou bugs em trechos de código \\
\hline
Sugestão de trechos de código durante a implementação de novas funcionalidades \\
\hline
Redução do tempo necessário para solucionar dúvidas técnicas pontuais durante a implementação \\
\hline
Auxílio na revisão de código (code review), identificando potenciais problemas ou inconsistências \\
\hline
Apoio na documentação de funções, classes e módulos diretamente no código \\
\hline
Facilidade em adaptar o código para diferentes linguagens de programação ou frameworks \\
\hline
Outro: \\
\hline
\end{tabular}
\label{tab:questoes-llms3}
\end{table*}

\begin{table*}[!ht].
\centering
\caption{Desvantagens percebidas no uso de LLMs em Engenharia de Software}
\begin{tabular}{|p{15cm}|}
\hline
\footnotesize
\textbf{Principais  Desvantagens identificadas} \\
\hline
Falta de transparência sobre a fonte ou fundamentação das respostas/sugestões fornecidas pelo modelo \\
\hline
Aumento do retrabalho devido à necessidade de revisar e corrigir o código gerado pelo LLM \\
\hline
Dependência excessiva do modelo, levando à diminuição do pensamento crítico e da autonomia dos desenvolvedores \\
\hline
Risco de vazamento de informações confidenciais ao utilizar LLMs hospedados externamente \\
\hline
Redução no aprendizado prático dos desenvolvedores mais iniciantes \\
\hline
Possibilidade de introdução de bugs devido a sugestões automáticas não totalmente compreendidas pelo desenvolvedor \\
\hline
Dificuldade em revisar ou manter código gerado automaticamente sem pleno entendimento de sua lógica \\
\hline
Sugestões desatualizadas, incompatíveis ou obsoletas para bibliotecas, frameworks ou padrões recentes \\
\hline
Risco de geração de código com falhas de segurança \\
\hline
Sugestões de código que não seguem os padrões ou práticas adotados pela equipe/projeto \\
\hline
Dificuldade em adaptar sugestões de código para contextos muito específicos ou altamente customizados \\
\hline
Geração de código excessivamente genérico, pouco otimizado ou fora do contexto do projeto \\
\hline
Dificuldade de integração do LLM com ferramentas e fluxos de trabalho já estabelecidos na equipe \\
\hline
Possíveis problemas éticos ou legais relacionados a direitos autorais do código sugerido \\
\hline
Outro: \\
\hline
\end{tabular}
\label{tab:questoes-llms4}
\end{table*}

\begin{table*}[!ht].
\centering
\caption{Desvantagens percebidas no uso de LLMs em Engenharia de Software}
\begin{tabular}{|p{15cm}|}
\hline
\footnotesize
\textbf{Questão aberta} \\
\hline
Qual sua opinião sobre o uso de LLMs na Engenharia de Software? \\
\hline
Como você tem utilizado LLMs na sua rotina? \\
\hline
\end{tabular}
\label{tab:questoes-llms5}
\end{table*}



\end{document}